\documentclass[preprint,aps]{revtex4}
\usepackage{amssymb}
\usepackage{amsmath}

\usepackage{graphicx}
\usepackage{dcolumn}
\usepackage{bm}
\usepackage{multirow}

\begin{document}

\begin{titlepage}

\title{Anomalous Dirac Plasmons in 1D Topological Electrides} 

\author{Jianfeng Wang$^1$, Xuelei Sui$^{2,1}$, Shiwu Gao$^1$, Wenhui Duan$^{2,4}$, Feng Liu$^{3,4}$\footnote{Email: fliu@eng.utah.edu}, and Bing Huang$^1$\footnote{Email: bing.huang@csrc.ac.cn}}

\address{$^1$ Beijing Computational Science Research Center, Beijing 100193, China}
\address{$^2$ Department of Physics and State Key Laboratory of Low-Dimensional Quantum Physics, Tsinghua University, Beijing 100084, China}
\address{$^3$ Department of Materials Science and Engineering, University of Utah, Salt Lake City, Utah 84112, USA}
\address{$^4$ Collaborative Innovation Center of Quantum Matter, Beijing 100084, China}

\date{\today}

\begin{abstract}

Plasmon opens up the possibility to efficiently couple light and matter at sub-wavelength scales. In general, the plasmon frequency is dependent of carrier density. This dependency, however, renders fundamentally a weak plasmon intensity at low frequency, especially for Dirac plasmon (DP) widely studied in graphene. Here we demonstrate a new type of DP, excited by a Dirac nodal-surface state, which exhibits an anomalously \emph{density-independent} frequency. Remarkably, we predict realization of anomalous DP (ADP) in 1D topological electrides, such as Ba$_3$CrN$_3$ and Sr$_3$CrN$_3$, by first-principles calculations. The ADPs in both systems have a density-independent frequency and high intensity, and their frequency can be tuned from terahertz to mid-infrared by changing the excitation direction. Furthermore, the intrinsic weak electron-phonon coupling of anionic electrons in electrides affords an added advantage of ultra-low phonon-assisted damping and hence a long lifetime of the ADPs. Our work paves the way to developing novel plasmonic and optoelectronic devices by combining topological physics with electride materials.

\end{abstract}

\maketitle

 \draft

\vspace{2mm}

\end{titlepage}

Plasmon, resulting from collective electron density oscillations due to long-range Coulomb interaction, dominates the long-wavelength elementary excitation spectrum in metals and doped semiconductors \cite{plasmon1}. It affords an important mechanism to efficiently couple light and matter at sub-wavelength scales, opening up the possibility to manipulate electromagnetic energy in nanophotonic and nano-optoelectronic devices \cite{plasmon1,plasmon2,plasmon3,plasmon4,plasmon5,plasmon6,plasmon7}. Recently, Dirac plasmons (DPs), as prototyped in graphene, have attracted great interest, because of their tunable frequency, enhanced light confinement and probable long lifetime \cite{graphpla1,graphpla2,graphpla3}. The typical terahertz (THz) or infrared (IR) response of a DP gives rise to promising applications in spectroscopy, biosensing, and security-related areas \cite{graphpla1,graphpla2,graphpla3,graphpla4,graphpla5,graphpla6,graphpla7}.

To date, all the known plasmons have a common feature: their frequency strongly depends on carrier density ($n$). For example, the long-wavelength plasmon frequency $\omega$ in metals and graphene follows the $n^{1/2}$ and $n^{1/4}$ power-law scaling \cite{graphpla3,graphpla4}, respectively. In addition, the plasmon intensity is proportional to the density of states (DOS) near the Fermi level ($E_F$), which is also related to $n$. These $n$-dependent features render the conventional DPs to have a weak intensity at low frequency, as illustrated by green line in Fig. 1(a). It poses a fundamental challenge to achieve a strong DP at THz/IR frequency.

In this Letter, we demonstrate an anomalous DP (ADP), excited by a unique Dirac nodal-surface (DNS) state, which exhibits a surprising \emph{$n$-independent} frequency and a constant high intensity. Most remarkably, we predict the existence of ADP in 1D topological electrides, where the DNS states are formed by anionic electrons trapped in the cavities. In addition, the loosely bound anionic electrons in electrides results in an ultra-weak phonon scattering and hence a long lifetime of the ADPs. Importantly, we further discover that Ba$_3$CrN$_3$ and Sr$_3$CrN$_3$ are ideal 1D topological electrides to realize the ADPs, having a tunable frequency from THz to mid-IR and an ultra-low phonon-assisted damping for a very long lifetime.

The classical long-wavelength plasmon frequency in 3D metals is written as $\hbar\omega_{\rm PM}=(4\pi e^2 n/\kappa m)^{1/2}$, where $m$ and $\kappa$ is the effective mass and dielectric constant, respectively. Recently, the emerging topological semimetal (TSM) states \cite{tsm1,tsm2,tsm3,tsm4} in solids, as characterized by a linear band crossing between conduction and valence bands in higher dimensions (i.e., 1D and 2D), provide a new platform to create novel DPs beyond graphene \cite{diracpla,nodelinepla1,nodelinepla2,diracplaexp}. For a TSM, the ``relativistic'' effective mass of quasiparticles is $n$-dependent, as $m_r \propto n^{1/3}$ and $n^{1/2}$ for Dirac nodal-point (DNP) and Dirac nodal-line (DNL) semimetals, respectively \cite{SM}. Consequently, the $n^{1/2}$ scaling in the classical plasmon frequency is partially offset by the $m_r$, resulting in $\omega_{\rm DNP}\sim n^{1/3}$ and $\omega_{\rm DNL}\sim n^{1/4}$ for DNP and DNL semimetals, respectively \cite{diracpla,nodelinepla1,nodelinepla2}. Interestingly, we realize when the dimensionality of band crossing is further increased from 1D DNL to 2D DNS \cite{dns1,dns2}, $m_r$ becomes proportional to $n$, $m_r\propto n$, to completely offset the $n^{1/2}$ scaling in the frequency. This means that the plasmon frequency of a DNS semimetal becomes independent of $n$, which is fundamentally different from all the known plasmons.

\begin{figure}[ht]
\centerline{ \includegraphics[clip,width=0.75\linewidth,angle=0]{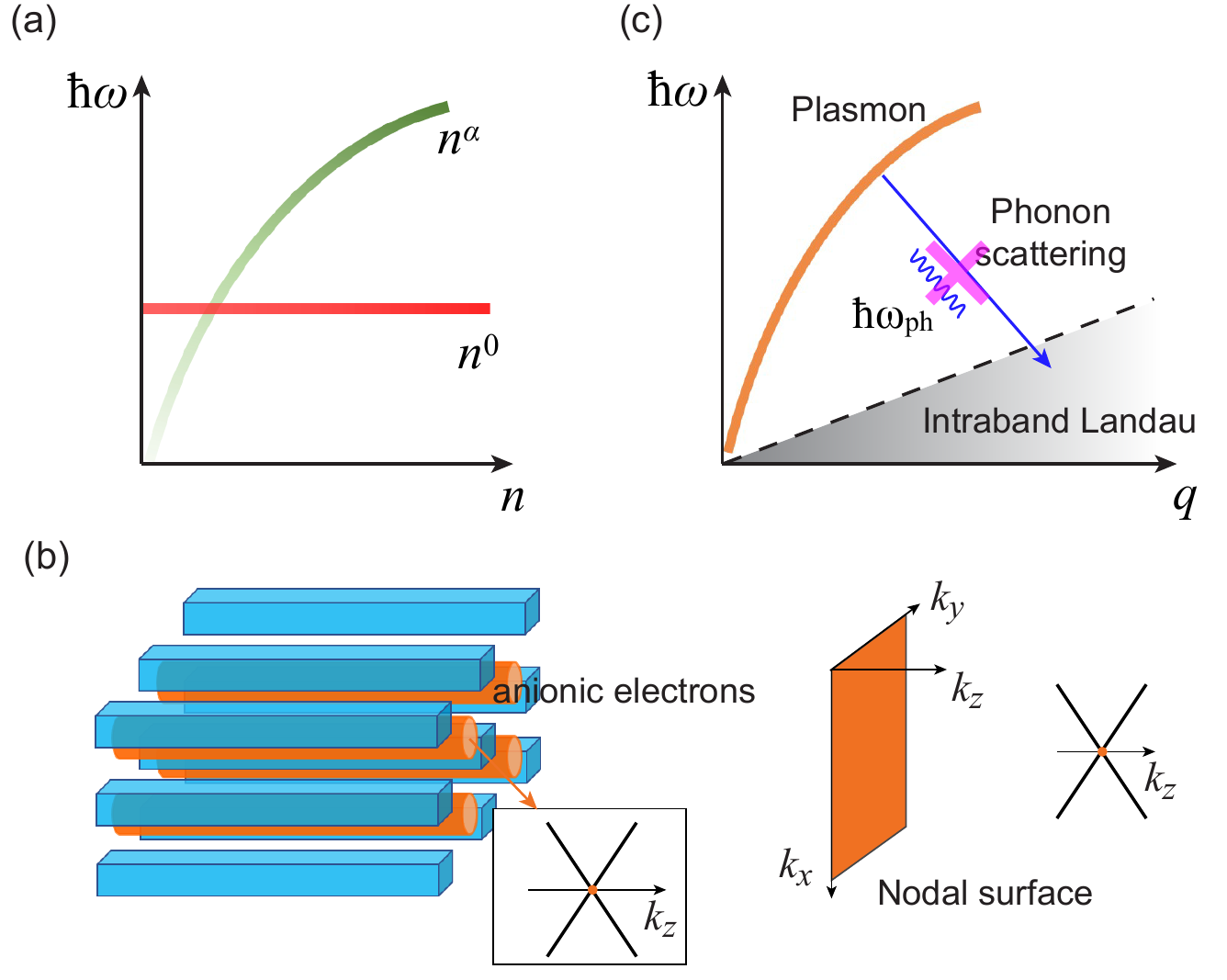}}
         \caption{(a) Density dependences of plasmon frequency. Green curve: the usual plasmon following a $\omega\sim n^{\alpha}$ scaling ($\alpha=1/2, 1/3, 1/4$ for parabolic metals, DNP, and DNL semimetals, respectively); Red curve: our proposed ADP following the $\omega\sim n^0$ scaling. The brightness of the color indicates the intensity of plasmon. (b) Left panel: schematic plot for 1D electride, where the orange channels denote the anionic electrons. Inset shows a band crossing of a single anionic electron chain along the kz direction. Right panel: The crossing points form a 2D degenerate DNS. (c) Schematic illustration of the phonon-assisted damping for Dirac plasmon. The plasmon may enter the intraband Landau region and decay into electron-hole pairs via phonon scattering or emitting a phonon. This damping pathway will be naturally minimized in electrides.
         }
\label{fig:1}
\end{figure}

To show this vigorously, based on random phase approximation (RPA) and irreducible polarizability of Dirac systems \cite{diracpla}, one can deduce the long-wavelength plasmon frequency for DNS state as \cite{SM}
\begin{equation}
\hbar \omega_{\rm DNS}=\sqrt{\frac{ge^2S\hbar v_F \cos^2\theta}{\pi^2\kappa}} + O(q^2),\label{eq:one}
\end{equation}
where $g$ is the degeneracy factor, $S$ is the DNS area in the momentum space, $v_F$ is the Fermi velocity along the normal direction of DNS, and $\theta$ is the angle between $\bm{q}$ and the normal direction of DNS. As shown by the red line in Fig. 1(a), the long-wavelength plasmon of DNS follows exactly a $\omega_{\rm DNS}\sim n^0$ scaling. Moreover, the 1D nature of DNS leads to a constant high DOS near $E_F$ \cite{SM}, which produces a constant strong plasmon intensity. Consequently, an ADP has inherently a $n$-independent frequency and high intensity. As indicated by Eq. (1), the frequency of ADP depends solely on the direction of plasmon excitation (a $|\cos\theta|$ function), providing a simple way to continuously tune its frequency.

Though the DNS states were proposed theoretically \cite{dns1,dns2}, they have not been observed in experiments due to the lack of ideal materials. Here we predict that the DNS states and ADPs can be realized in 1D topological electrides. Electrides are known as special ionic solids, in which excess electrons are trapped in the cavities to serve as anions \cite{electride1,electride2,electride3,electride4}. Generally, they can be classified into 2D, 1D and 0D electrides according to the dimensionality of the confinement \cite{electride5}. The anionic electrons have a low work function, because they are rather loosely bound to atoms or lattice. As a result, they may appear as occupied states near $E_F$, suitable for band inversion or band crossing \cite{topoelectride}. As shown in Fig. 1(b), when the anionic electrons are confined in 1D channels (e.g., $z$ direction), they might generate a band crossing along $k_z$ direction in the momentum space. Their inter-channel coupling in periodic array translates the crossing points along the $k_x$ and $k_y$ directions in the 3D Brillouin zone (BZ). The resulting 2D band crossing can be protected by some specific crystal symmetries to realize a DNS state. In addition, differing from the conventional topological materials, there is a negligible spin-orbit coupling (SOC) effect for loosely bound anionic electrons, which is also vital for achieving the spinless nodal states.

We note that the nature of electrides also naturally makes the excited ADP to have a low damping. In general, there are several possible damping pathways for plasmons \cite{damping1}, including direct decay into electron-hole pairs via intra- or inter-band Landau damping, scattering from impurities or defects, and inelastic scattering with phonons. For DPs, the first two processes can be effectively reduced by changing $n$ and improving sample quality, respectively. Thus, the phonon-assisted damping is usually the dominant loss pathway \cite{damping1,damping2,damping3}. As shown in Fig. 1(c), a DP can decay into an electron-hole pair via emission of a phonon. Here for our proposed ADPs in electrides, the loosely bound anionic electrons generally exhibit a weak electron-phonon (e-ph) coupling with a negligible phonon scattering \cite{Ca2Ne-ph}. Consequently, the phonon-assisted damping pathway can be largely suppressed.

The above analysis indicates that an ADP generated by a DNS state in 1D electrides could overcome both low intensity and phonon-assisted damping associated with the conventional DPs. Based on an extensive structural search in the ICSD database, we successfully identify that Ba$_3$CrN$_3$ and Sr$_3$CrN$_3$ are ideal 1D electrides to realize the desired ADP.

\begin{figure}[ht]
\centerline{ \includegraphics[clip,width=0.8\linewidth,angle=0]{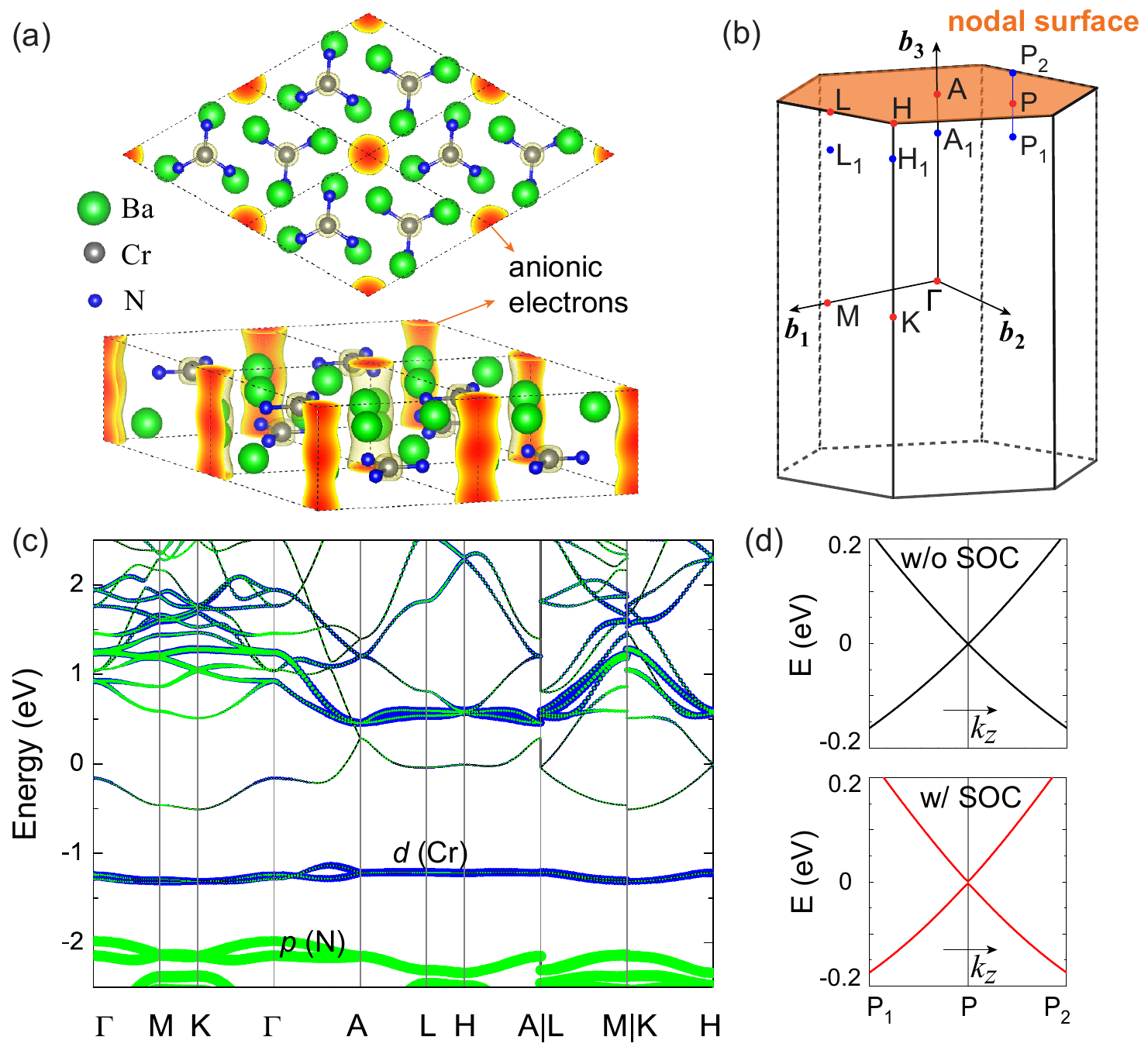}}
         \caption{(a) Top (upper panel) and side (lower) view of crystal structure of Ba$_3$CrN$_3$. The red regions indicate partial charge density distribution around $E_F$ ($-0.3$ to $0.3$ eV). (b) The first BZ of Ba$_3$CrN$_3$. The high-symmetry and some arbitrary points (dots) are labelled. The nodal surface is formed at the $k_z = \pi/c$ plane (orange). (c) Band structure with atomic orbital projections. The Fermi level is set to zero. (d) Magnified band structure along an arbitrary line $P_1$--$P$--$P_2$ perpendicular to the $k_z = \pi/c$ plane at an arbitrary $P$ point located in the plane, as indicated in (b).
         }
\label{fig:2}
\end{figure}

Ba$_3$CrN$_3$ and Sr$_3$CrN$_3$ have already been synthesized in experiments \cite{expstruc}. Here we take Ba$_3$CrN$_3$ as an example to illustrate its ADP excitations, and leave the results of Sr$_3$CrN$_3$ in the Supplementary Information \cite{SM}. As shown in Fig. 2(a), the unit cell of hexagonal Ba$_3$CrN$_3$ (space group $P6_3/m$) contains two trigonal planar CrN$_3$ anions that are related to each other by an inversion or screw rotation symmetry \cite{expstruc}. Along the $z$ direction, a 1D cavity is formed at the corner of the unit cell. Our calculations indicate that Ba$_3$CrN$_3$ is nonmagnetic with zero magnetic moment on Cr atoms. This is because Cr has a $+4$ charge state, and its remaining two $d$ electrons fill a singlet state. Thus, there are two ``excess'' electrons remaining per unit cell to act as ``anions''. The band structure along high-symmetry lines in BZ [Fig. 2(b)] is shown in Fig. 2(c), which indicates a semi-metallic feature of Ba$_3$CrN$_3$. Strikingly, the two energy bands near $E_F$ are mostly contributed by the anionic electrons with negligible atomic orbital components. The band dispersions are small in the $k_z = 0$ and $k_z = \pi/c$ planes but large along the $k_z$ direction, indicating a 1D nature. The partial charge densities of these two bands around $E_F$ further show that the anionic electrons are confined in the channels of 1D cavity [Fig. 2(a)]. Hence, Ba$_3$CrN$_3$ is a 1D electride with an effective formula [Ba$_6$Cr$_2$N$_6$]$^{2+}\cdot 2e^{-}$.

Figure 2(c) also shows that the conduction and valence bands are degenerate along the high-symmetry paths $A$--$L$--$H$--$A$, but split along $\Gamma$--$A$, $M$--$L$, and $K$--$H$. Actually, such band degeneracy occurs at all points in the $k_z = \pi/c$ plane, as confirmed by plotting the bands [Fig. 2(d)] along an arbitrary line $P_1$--$P$--$P_2$ perpendicular to the $kz = \pi/c$ plane at an arbitrary $P$ point in the plane [see Fig. 2(b)]. So, the band crossing takes place throughout the BZ boundary to form a perfect DNS state [orange plane in Fig. 2(b)].

The DNS state is protected by a nonsymmorphic symmetry \cite{dns1}. Ba$_3$CrN$_3$ has time-reversal symmetry $T = K$, with $K$ being the complex conjugation, inversion symmetry $I$, and screw rotation symmetry $S_z =\{C_{2z}|c/2\}$, combining a twofold rotation and a half-lattice translation along the $c$ axis. Two compound symmetries, $IT$ and $IS_z$, are preserved in the $k_z = \pi/c$ plane, and their anticommutation ensures the twofold band degeneracy in the entire $k_z = \pi/c$ plane \cite{dns1}. When SOC effect is included, this degeneracy of DNS is lifted but the SOC gap is negligibly small due to the unique nature of anionic electrons [see Fig. 2(d)]. The similar conclusions are drawn for Sr$_3$CrN$_3$ \cite{SM}.

\begin{figure}[ht]
\centerline{ \includegraphics[clip,width=0.75\linewidth,angle=0]{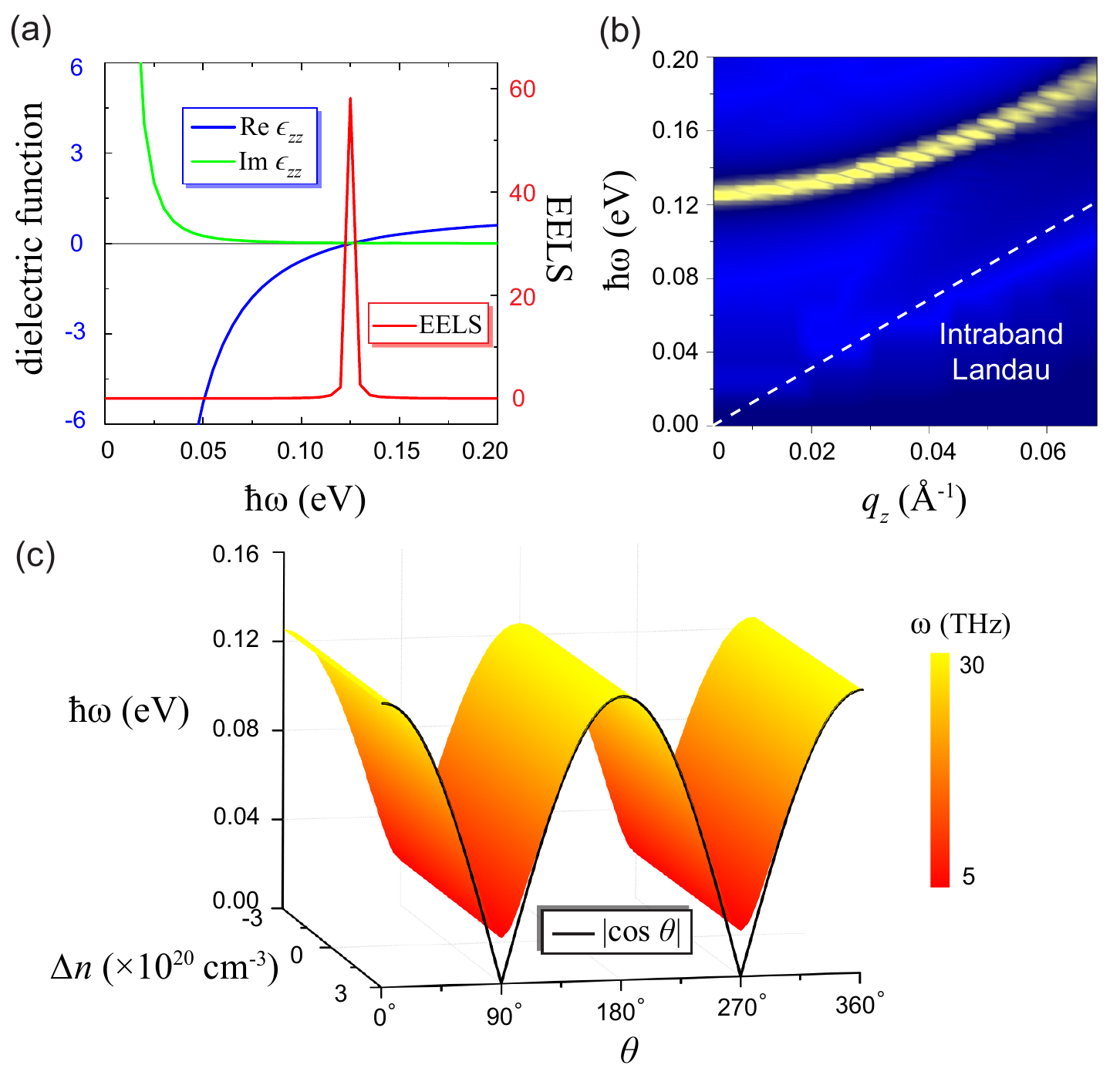}}
         \caption{(a) Real and imaginary parts of dielectric function and EELS of Ba$_3$CrN$_3$ as functions of frequency with $q$ = 0.001 {\AA}$^{-1}$ along the $z$ direction. (b) EELS as a function of frequency $\hbar\omega$ and wave vector $q_z$. The white dashed line denotes the upper edge $\hbar v_F q$ of intraband particle-hole continuum. (c) Plasmon frequency as a function of $\Delta n$ and excitation direction ($\theta$). The color from red to yellow indicates the frequency from THz to mid-IR, and the brightness of color indicates almost a constant high plasmon intensity.
         }
\label{fig:3}
\end{figure}

After establishing an ideal DNS state in Ba$_3$CrN$_3$, we set to investigate its plasmonic excitations. Under RPA, the collective plasmon mode can be determined by the dynamical dielectric function $\epsilon(\bm{q},\omega)=1-V(q)\Pi(\bm{q},\omega)$, where $V(q)=4\pi e^2/\kappa q^2$ is the Coulomb interaction in the wave vector space, and $\Pi(\bm{q},\omega)$ is the irreducible polarizability function. In the long-wavelength limit ($q\rightarrow 0$), a noninteracting irreducible polarizability is given by \cite{diracpla,Siplasmon}
\begin{eqnarray}
\Pi(\bm{q},\omega)=-\frac{2}{(2\pi)^3}\int d^3\bm{k}\sum_{l,l'}|\langle \bm{k}+\bm{q},l'|e^{i\bm{q}\cdot\bm{r}}|\bm{k},l\rangle|^2
\times \frac{n_F(E_{\bm{k},l})-n_F(E_{\bm{k}+\bm{q},l'})}{\hbar\omega+E_{\bm{k},l}-E_{\bm{k}+\bm{q},l'}+i\eta}, \label{eq:two}
\end{eqnarray}
where $n_F$ is the Fermi-Dirac distribution function. The collective plasmon mode is defined at zeros of the complex dynamical dielectric function. It is more convenient to calculate the electron energy loss spectrum (EELS), i.e., EELS=$-$Im$[1/\epsilon(\bm{q},\omega)]$, whose broadened peaks indicate the plasmons \cite{Siplasmon}. Along the $z$ direction, we calculate the long-wavelength ($q$ = 0.001 {\AA}$^{-1}$) dielectric function of Ba$_3$CrN$_3$ and EELS at $T =$ 300 K, as shown in Fig. 3(a). It can be seen that a sharp plasmon peak with high intensity appears for Re$[\epsilon(\bm{q},\omega)]=0$. Simultaneously, the Im$[\epsilon(\bm{q},\omega)]\rightarrow 0$, indicating a weak direct damping rate. The plasmon excitation energy of 0.125 eV corresponds to a frequency of $\sim 30$ THz in the mid-IR range. As shown in Fig. 3(b), the calculated plasmon dispersion of Ba$_3$CrN$_3$ over a small range of wave vector has a normal $q^2$ relationship because of a 3D dielectric screening. Importantly, the plasmon excitations lie above the region of intraband Landau damping, consistent with the sharp peak of plasmon; it indicates that the direct decay into electron-hole pairs is almost forbidden.

In Fig. 3(c), we plot the plasmon frequency as a function of density $n$ and excitation direction angle $\theta$ at the long-wavelength limit. Remarkably, at any fixed excitation direction, the plasmon frequency keeps a constant value ($\omega\sim n^0$) for a significantly large range of changing density up to $\sim 3\times10^{20}$ cm$^{-3}$ (i.e., within the energy range of linear band dispersion \cite{SM}), consistent with our predicted feature of ADP. Meanwhile, a high intensity of ADP remains almost unchanged for different $n$ \cite{SM}, because of a constant DOS within the same energy range. On the other hand, the ADP frequency depends on plasmon direction $\theta$ and can be continuously tuned, following a $|\cos\theta|$ function [black curve in Fig. 3(c)], which agrees well with our model [Eq. (1)]. It is noted that the plasmon mode along the $x$ direction (i.e., $\theta=90^{\circ}$) has a small excitation energy ($\sim 20$ meV) and enters the particle-hole continuum at a large $q$ \cite{SM}. While the non-zero frequency for $\theta=90^{\circ}$ is due to the fact that the entire DNS band has a small dispersion in the $xy$ plane. Moreover, the high intensity of ADP almost maintains for different plasmon directions, as confirmed by our calculations \cite{SM}. As a result, by changing $\theta$, a high-intensity ADP with a frequency tuned from THz to mid-IR can be achieved [Fig. 3(c)]. In addition, we find the plasmon frequency is robust against strain, due to the nature of loosely bound anionic electrons \cite{SM}.

\begin{figure}[ht]
\centerline{ \includegraphics[clip,width=0.8\linewidth,angle=0]{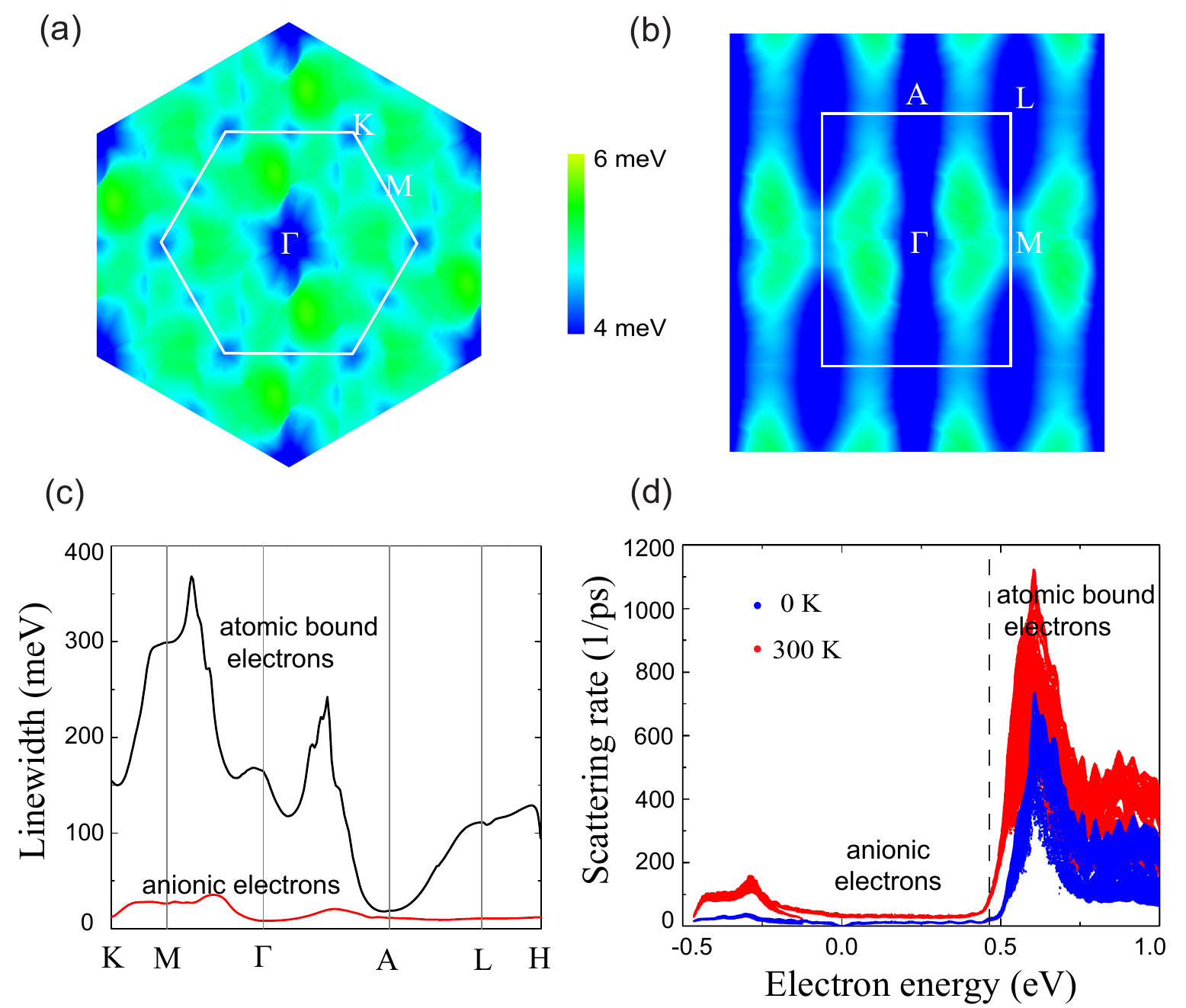}}
         \caption{Average e-ph coupling matrix element $|g_{mn,\nu}(\bm{k},\bm{q})|$ of all 42 phonon modes for the DNS band of Ba$_3$CrN$_3$ at the $A$ point of BZ as a function of phonon wave vector $\bm{q}$ in the (a) $q_z=0$ and (b) $q_x=0$ plane. (c) Electron linewidth for the DNS band of anionic electrons (red line) and for the higher band of atomic bound electrons (black line) along high-symmetry $\bm{k}$ points with a converged smearing value of $\eta=10$ meV. (d) Scattering rate of electron as a function of energy at 0 K (blue dots) and 300 K (red dots), obtained with 64000 random $\bm{k}$-points and $\bm{q}$-points respectively. The Fermi level is set to zero.
         }
\label{fig:4}
\end{figure}

Finally, we discuss the phonon scattering of ADPs in Ba$_3$CrN$_3$, by calculating the e-ph interaction for the DNS states. The electron linewidth or scattering rate is directly connected to the imaginary part of the e-ph self-energy \cite{e-ph1}
\begin{equation}
\begin{aligned}
\Sigma_{n,\bm{k}}''=&\pi\sum_{m\nu \bm{q}}|g_{nm\nu}(\bm{k},\bm{q})|^2 \times [(N_{\nu\bm{q}}^0+f_{m\bm{k}+\bm{q}}^0)\delta(\varepsilon_{n\bm{k}}-\varepsilon_{m\bm{k}+\bm{q}}+\hbar\omega_{\nu\bm{q}}) \\
&+(N_{\nu\bm{q}}^0+1-f_{m\bm{k}+\bm{q}}^0)\delta(\varepsilon_{n\bm{k}}-\varepsilon_{m\bm{k}+\bm{q}}-\hbar\omega_{\nu\bm{q}})], \label{eq:three}
\end{aligned}
\end{equation}
where $\hbar\omega_{\nu\bm{q}}$ is the phonon energy, $N_{\nu\bm{q}}^0$ ($f_{m\bm{k}+\bm{q}}^0$) is the Bose-Einstein (Fermi-Dirac) distribution, and $g_{nm\nu}(\bm{k},\bm{q})$ is the e-ph coupling matrix element corresponding to electron scattering from band $n$ at $\bm{k}$ to band $m$ at $\bm{k}+\bm{q}$ by phonon $\nu$ with $\bm{q}$. Here, a Wannier-Fourier interpolation method \cite{epw} is used to obtain the numerical results of e-ph coupling.

In Figs. 4(a) and (b), we plot $\bm{q}$-dependent average e-ph coupling matrix element of all 42 phonon modes for electron scattering from the DNS band at the $A$ point of BZ. It is found that the e-ph interactions are indeed ultra-weak and the average coupling matrix element is about several meV, which is two orders of magnitude smaller than that of graphene \cite{graphe-ph}. Similar results are obtained at the other momentum points \cite{SM}. The calculated electron linewidths for the occupied DNS band of anionic electrons and for an unoccupied band of atomic bound electrons are shown in Fig. 4(c). Remarkably, the former is much smaller than the latter over the entire BZ. Correspondingly, the scattering rate of anionic electrons at low energy is much lower than that of atomic bound electrons at high energy, as shown in Fig. 4(d). As a result, the relaxation time is very long (e.g., about 30 fs at $T=$300 K) for the DNS states, even longer than that of graphene plasmons \cite{damping2}. According to Eq. (3), the scattering rate is related to both the e-ph coupling matrix elements and the DOS. The low electron scattering of graphene is mostly contributed by the low DOS near $E_F$, while the carrier doping will inevitably result in a rapid increase of scattering rate \cite{graphe-ph}. In contrast, the ultra-weak e-ph coupling and nearly constant DOS exist in a wide energy range for Ba$_3$CrN$_3$, which leads to an ultra-low phonon scattering rate over a large range of doping [Fig. 4(d)].

In conclusion, we demonstrate an ADP realized by exciting a DNS state in 1D electride, as exemplified in Ba$_3$CrN$_3$ and Sr$_3$CrN$_3$. It exhibits an anomalous density-independent frequency, which enables a low-frequency plasmon with high intensity. Furthermore, the weak e-ph interaction in electride materials reduces the phonon-assisted damping of ADP, leading to long-lived plasmons. Our work paves the way to developing plasmonic applications by combining topological states with electride materials, which are expected to draw broad interests especially from experimentalists.

J. Wang and X. Sui contributed equally to this work. The authors thank L. Kang and X. Zhang for helpful discussions. J.W. and B.H. acknowledge the support from NSFC (Grant No. 11574024) and NSAF U1530401. X.S. and W.D. acknowledge support from MOST of China (Grant No. 2016YFA0301001) and NSFC (Grants No. 11674188 and No. 11334006). F.L. acknowledge the support from US-DOE (Grant No. DE-FG02-04ER46148). Part of the calculations were performed at Tianhe2-JK at CSRC.

\end{document}